\documentclass[aps,prx,preprint,bibliography,superscriptaddress]{revtex4-1}
\usepackage[utf8]{inputenc}
\usepackage{amsmath}
\usepackage{physics}
\usepackage{mathtools}
\usepackage{amsfonts}
\usepackage{amssymb}
\usepackage{braket}
\usepackage{graphicx}
\usepackage{multirow}
\usepackage{subfigure}
\usepackage{appendix}
\usepackage{comment}
\usepackage{textcomp}
\usepackage{color}
\usepackage[dvipsnames]{xcolor}
\usepackage{mathrsfs}
\usepackage{natbib}

\begin{document}
\title{Uncovering Exceptional Contours in non-Hermitian Hyperbolic Lattices}
\author{Nisarg Chadha}
\affiliation{Undergraduate Programme, Indian Institute of Science, Bangalore 560012, India}
\author{Awadhesh Narayan}
\email{awadhesh@iisc.ac.in}
\affiliation{Solid State and Structural Chemistry Unit, Indian Institute of Science, Bangalore 560012, India}

\date{\today}

\begin{abstract}
Hyperbolic lattices are starting to be explored in search of novel phases of matter. At the same time, non-Hermitian physics has come to the forefront in photonic, optical, phononic, and condensed matter systems. In this work, we introduce non-Hermitian hyperbolic lattices and elucidate its exceptional properties in depth. We use hyperbolic Bloch theory to investigate band structures of hyperbolic lattices in the presence of non-Hermitian on-site gain and loss as well as non-reciprocal hopping. Using various analytical and numerical approaches we demonstrate widely accessible and tunable exceptional points and contours in \{10,5\} tessellations, which we characterize using phase rigidity, energy scaling, and vorticity. We further demonstrate the occurrence of higher-order exceptional points and contours in the \{8,4\} tessellations using the method of Newton polygons, supported by vorticity and phase rigidity computations. Finally, we investigate the open boundary spectra and densities of states to compare with results from band theory, along with a demonstration of boundary localization. Our results unveil an abundance of exceptional degeneracies in hyperbolic non-Hermitian matter.
\end{abstract}

\maketitle

\section{Introduction}

Spaces with negative curvature emerge naturally in general relativity~\cite{friedmann1979curvature}, and find applications in graph theory~\cite{brooks1996some}, random walks~\cite{monthus1996random}, complexity theory~\cite{krioukov2010hyperbolic}, and quantum information theory~\cite{breuckmann2016constructions, breuckmann2017hyperbolic}. The enhanced bulk connectivity of the system~\cite{mertens2017percolation} makes hyperbolic surfaces an efficient candidate for data storage and communication, making these geometries ubiquitous in data science and electrical engineering. 
Recent works on networks of coupled microwave resonators and superconducting qubits~\cite{kollar2019hyperbolic, chen2023hyperbolic} provide more tangible insight into these ideas from the perspective of band theory.

The theoretical extension of ideas from band theory and topology to hyperbolic geometries~\cite{maciejko2021hyperbolic} and their experimental realisations~\cite{kollar2019hyperbolic,chen2023hyperbolic} on tabletop experimental platforms have propelled hyperbolic lattices to the forefront in the search for novel phases of matter. 
Progress in circuit quantum electrodynamics has made it possible to simulate hyperbolic lattices and explore band theory on these models, therefore being a direct probe of the influence of the geometry and metric on the properties of the system. 
This has motivated the study of hyperbolic lattices, which are regular tilings of hyperbolic space.
Tilings of regular $p$-gons with coordination number $q$ are denoted in the Schläfli notation as \{$p$,$q$\}.

The negative curvature of hyperbolic surfaces endows interesting properties to these systems, such as the finite ratio of boundary sites to the total sites and the non-Abelian nature of the translation group.

\begin{figure}
     \includegraphics[width=\textwidth]{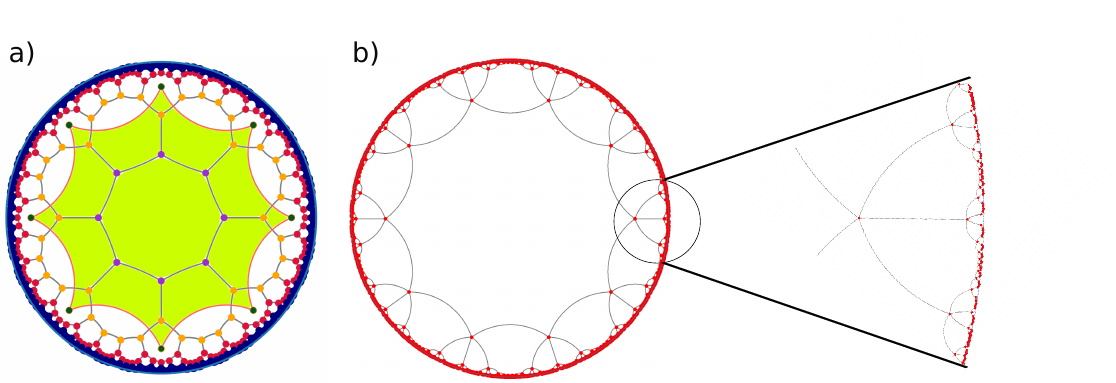}
     \caption{\textbf{Visualising the hyperbolic plane in two dimensions through the Poincar\'e disk model.} (a) The \{8,3\} tessellation with different coloured sites denoting different epochs generated recursively. The area in green shows the 16-site unit cell, also called the Bolza lattice, which itself is an \{8,8\} tessellation. (b) The \{10,5\} tessellation up to the first four epochs, along with a zoomed-in image showing the sites near the boundary.} \label{fig:Schematic}
\end{figure}

Recent works have investigated hyperbolic analogues of topological insulators~\cite{urwyler2022hyperbolic, liu2022chern}.
The use of real-space topological invariants on flat projections of hyperbolic tessellations has shown Chern insulator phases accompanied by topological edge modes and a quantized conductance~\cite{liu2022chern}. 
Topological Anderson insulator phases have also been shown to arise in the presence of symmetry-preserving disorder. 
Furthermore, higher-order topological insulator phases have been discovered in 2D hyperbolic tessellations~\cite{liu2023higher,tao2023higher}. 
These phases show zero-dimensional corner edge states whose degeneracy depends on the symmetry of the crystal. 
The realisation of hyperbolic lattices with arbitrary rotational symmetries allows these corner edge states to have higher degeneracies than possible for Euclidean systems. 
The effects of flux threading through the hyperbolic lattice have also been studied~\cite{PhysRevB.106.155120, stegmaier2022universality}.
Remarkably, a universality in the shape of the Hofstadter butterfly is obtained when the unit cells in the compactified manifold are threaded by a flux~\cite{stegmaier2022universality}. 
These Hofstadter butterfly patterns turn out to be independent of the coordination number $q$ for tessellations of the same $p$-gons.

Unlike Euclidean lattices, hyperbolic lattices show a finite fraction of boundary sites in the thermodynamic limit.
Consequently, an open boundary analysis gets prominent contributions from the boundary sites.
As a result, open boundary spectra show considerable deviations from the bulk spectra even in the thermodynamic limit.
Refs.~\cite{maciejko2022automorphic,stegmaier2022universality,chen2023hyperbolic} have implemented a compactification of the hyperbolic manifold using regular maps to identify boundary sites to enforce periodic boundaries on a finite to obtain the true bulk spectra.

In a parallel development, non-Hermitian systems have been in the limelight in recent years due to their intriguing fundamental properties, with a potential for interesting applications~\cite{ashida2020non,el2018non,bergholtz2021exceptional,ghatak2019new,banerjee2022non}. 
Beginning with the pioneering ideas of Bender and co-workers~\cite{PhysRevLett.80.5243,bender2007making}, the interplay with topology has reinvigorated the field and has led to several seminal discoveries. 
The role of topology in non-Hermitian systems is now at the forefront of research, with implications for photonics, optics, metamaterials, topoelectric circuits and many others. 
Exceptional points (EPs) are an important feature of non-Hermitian systems. 
These are singularities where eigenvalues and eigenvectors become identical~\cite{heiss2012physics,miri2019exceptional,ozdemir2019parity}, and have no counterpart in Hermitian systems. 
The order $n$ of the EP is the number of eigenvectors coalescing together.
In this regard, higher-order EPs ($n>$2) have attracted interest~\cite{hodaei2017enhanced,chen2017exceptional,wiersig2020review} as higher-order branch points in the energy spectrum.
Not only are they interesting from a fundamental point of view -- with remarkable associated properties such as Berry phases, Riemann sheet structures, and bulk Fermi arcs -- they are also beginning to be exploited in applications. 
For instance, EPs have been used to design sensors with enhanced sensitivity~\cite{PhysRevLett.112.203901, wiersig2020review,de2022non}. 
Another intriguing phenomenon which has been recently identified in non-Hermitian systems is the non-Hermitian skin effect (NHSE), where a macroscopic fraction of states localize at the boundary~\cite{yao2018edge,alvarez2018non,yokomizo2019non,borgnia2020non,zhang2022review}. 
The NHSE is another topic of very active research in the past couple of years, whose implications are only beginning to be understood. 
It features intriguing connections to spectral topology, and a plethora of experimental platforms have been used to implement its phenomenology~\cite{zhang2022review,lin2023topological}.

In this work, we explore the interplay of non-Hermiticity and hyperbolic geometry by using analytic band theory as well as numerical tight-binding calculations. 
We find that the non-Abelian nature of the translational group leads to a greater degree of freedom in parameter space, allowing the tuning of parameters to readily obtain EPs as well as their higher dimensional analogues, exceptional contours. 
We investigate the behaviour of EPs in the presence of different kinds of non-Hermitian interactions and characterise their properties. 

The \{10,5\} tessellation lies in the same infinite crystalline family of $\{2(2g+1),2g+1\}$ tessellations (with $g$ being the genus) as the honeycomb lattice in Euclidean space, and has been termed "hyperbolic graphene" ~\cite{boettcher2022crystallography}. 
Just as in graphene, the \{10,5\} tessellation has a two-site unit cell, making it a convenient, analytically tractable two-band model to study the behaviour of exceptional contours.
However, the presence of only two bands precludes higher-order EPs.
To demonstrate the tunability of higher-order EPs, we use the \{8,4\} lattice, which has a four-site unit cell.
We see that this four-band system gives us a simple geometry to realize second as well as fourth order EPs.

The rest of this work is organised as follows:
We briefly introduce the model for projecting hyperbolic space onto Euclidean space in Sec.~\ref{sec:understanding}. 
In Sec.~\ref{sec:tbmodel}, we introduce the tight-binding model on the \{10,5\} lattice and discuss the spectra obtained for the Bloch Hamiltonian under different non-Hermitian perturbations.
Non-Hermiticity is introduced using gain and loss and non-reciprocal hoppings in the unit cell.
Sec.~\ref{sec:exdiag} discusses the EPs obtained for the non-Hermitian Hamiltonian.
Phase rigidity and vorticity (spectral winding) are used to characterize the behaviour of the exceptional contours by quantifying the coalescing of the eigenstates.
The \{8,4\} lattice with a four-site unit cell provides a tractable model to show the onset of higher(fourth)-order EPs in a hyperbolic lattice.
We also utilize the recently proposed method of Newton polygons~\cite{jaiswal2021characterizing} to obtain the conditions for higher-order EPs. 
In Sec.~\ref{sec:realspace}, we construct the real-space lattice and use exact diagonalisation to obtain the energy spectra and the densities of states. 
Deviations in the results obtained using hyperbolic band theory and exact diagonalisation are discussed.
We summarise our work in Sec.~\ref{sec:summary} and motivate experimental realizations of hyperbolic non-Hermitian lattices.

\section{Understanding Hyperbolic Geometry}\label{sec:understanding}

We briefly summarize here essential concepts in hyperbolic geometry as a foundation for the rest of our results. 
We will be interested in space-filling tilings of regular $p$-gons, with each vertex having a coordination number of $q$. 
This is denoted in the Schl\"{a}fli notation as a \{$p$,$q$\} tessellation. 
In this notation, the square lattice is a \{4,4\} tessellation, and similarly, the honeycomb and triangular lattices are the \{6,3\} and \{3,6\} tessellations, respectively. 
Due to the restriction imposed by the angle sum property, these are the only permissible tessellations for Euclidean space. 
The absence of this restriction allows hyperbolic systems to have an infinite number of realisations of \{$p$,$q$\} tessellations where $(p-2)(q-2)>4$. 

2D hyperbolic spaces are often represented using projections onto Euclidean 2D spaces endowed with a non-Euclidean metric.
One such model for projection is the Poincar\'e disk model.
This projects the hyperbolic space onto a unit disk with distances measured in the Poincar\'e metric and is useful for compact depictions of large lattice sizes as shown in Fig.~\ref{fig:Schematic} for the \{8,3\} and \{10,5\} tessellations.
We have discussed the procedure to generate the Poincaré model for hyperbolic lattices in Appendix~\ref{app:poinc}.

The lattice translation operators for hyperbolic lattices are elements of the discrete non-Abelian Fuchsian symmetry group\cite{yuncken2011regular}.
This marks a striking deviation from Euclidean lattices, where the translation operators commute, and leads to the presence of higher-dimensional irreducible representations(irreps) for the symmetry groups of hyperbolic lattices.

Since Abelian groups can only have 1D irreps, the eigenstates of Euclidean systems with translational symmetry transform with a U(1) phase under the action of the translation operators.
This is the statement of Bloch's theorem.
Clearly, we cannot use Bloch's theorem for an arbitrary hyperbolic lattice, since there will be eigenvectors which transform as higher-dimensional irreps of the symmetry group.
Nevertheless, automorphic band theory has been used for hyperbolic lattices to describe the eigenstates that transform as a U(1) irrep.

Thus, hyperbolic Bloch theory is expected to be incomplete due to the presence of eigenvectors which do not transform as 1D-irreps.
We shall subsequently visit the deviation between results from open boundary numerical diagonalisation and automorphic Bloch theory for different non-Hermitian parameters in Section \ref{sec:realspace}.

\begin{figure}
     \includegraphics[width=0.9\textwidth]{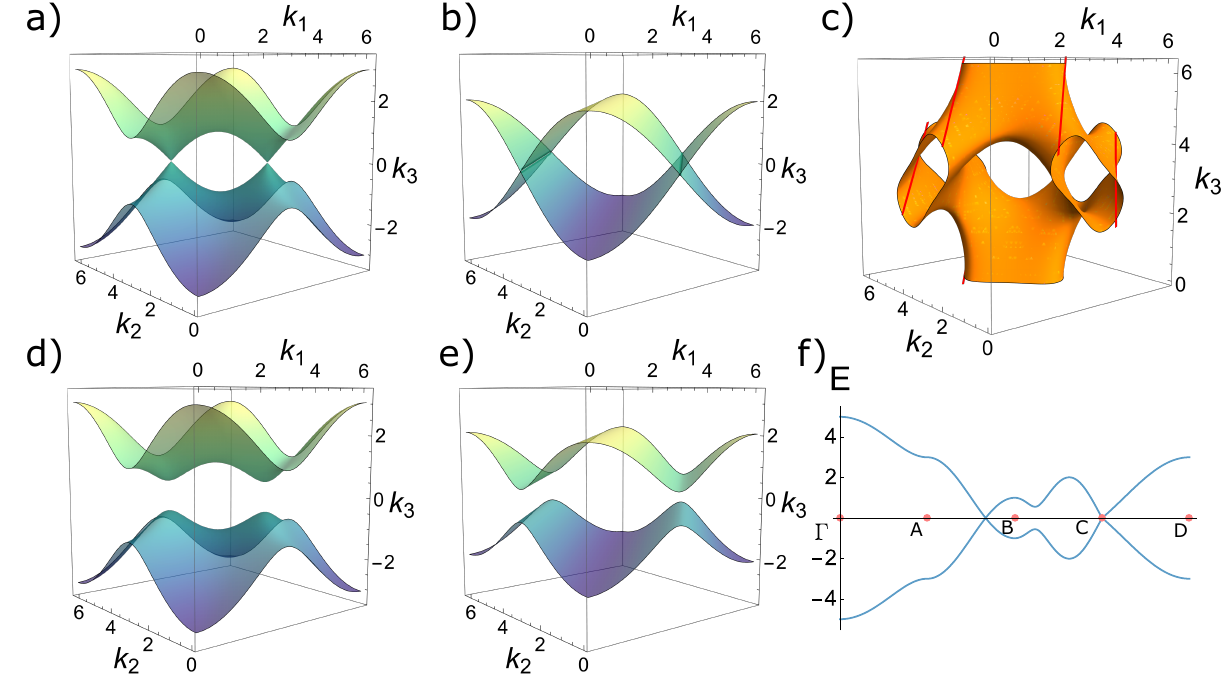}
     \caption{\textbf{Visualising the energy spectra for Hermitian hyperbolic graphene.} (a) The energy spectrum with $M=0$ produces the characteristic Dirac cones when $k_3=0$, $k_4=\pi$ in which case $h(\mathbf{k})=1+e^{ik_1}+e^{ik_2}$ becomes the phase factor for Euclidean graphene. (b) Nodal lines are obtained with $M=0$ for $k_3=2\pi/3$ and $k_4=4\pi/3$, and the condition for band touching becomes $k_1=k_2\pm\pi$. Modulating $k_3$, $k_4$ gives different shapes of nodal surfaces. (c) The surface represents the allowed values of $k_1,k_2,k_3$ at which there is a $k_4$ that produces a node. The lines shown in red along the surface are the values of $k_1,k_2,k_3$ where $k_4=0$ gives a node. (d) Adding an on-site potential ($M\neq 0$) opens a gap in the system. For $k_3=0$, $k_4=\pi$ a gapped spectrum ($\Delta E_g=2M$) is obtained. The linear dispersion near the extrema is replaced by a quadratic dispersion ($|E|=\frac{(\Delta k)^2}{2M}$). (e) For $k_3=2\pi/3$, $k_4=4\pi/3$, a gapped system is obtained with the linear scaling being replaced by a quadratic scaling as in (d). (f) Energy spectrum through the Brillouin zone. The points defined here are $\Gamma$(0,0,0,0), A$(\pi,0,0,0)$, B$(\pi,0,\pi,-\pi)$, C$(2\pi/3,-2\pi/3,0,\pi)$, and D$(\pi,\pi,\pi,\pi)$.}  
     \label{fig:Hermspectra}
\end{figure}

\section{The Hyperbolic Tight-Binding Model}\label{sec:tbmodel}

To implement the Bloch ansatz, we need to obtain the unit cell that can generate the entire lattice through the discrete translations.
The imposition of periodic boundary conditions leads to the identification of pairs of edges such that the lattice becomes a closed surface with only one face(such as the torus in 2D).
The generalization of the notion of a unit cell to hyperbolic lattices was carried out in Ref.~\cite{sausset2007periodic,boettcher2022crystallography} to obtain the conditions on \{$p$,$q$\} such that we can obtain a closed primitive unit cell by imposing periodic boundaries.
This leads to the condition that the identified unit cell must correspond to a surface with only one face, giving rise to constraints on $p$ and $q$.
Ref.~\cite{boettcher2022crystallography} detail the families of lattices($\{p,q\}$) whose Bravais lattices have a unit cell.

Of interest to us is the $\{2g+1,2(2g+1)\}$ family, where $g$ is the genus of the Brillouin zone.
$g=1$ gives us the honeycomb lattice, and $g=2$ gives the \{10,5\} lattice.
Like graphene(\{6,3\}), we obtain two sublattices, each forming a $\{2g+1,2(2g+1)\}$ tessellation of its own.
In Sec.~\ref{sec:higher}, we shall also use the \{8,4\} lattice, which is part of the $\{4g,4\}$ family with a $2g$ site unit cell.
 
To proceed with the hyperbolic Bloch ansatz in a tight binding model, we choose a unit cell and identify the nearest neighbours in terms of the translation operators. 
These have been worked out for the \{10,5\} tessellation in Ref.~\cite{chen2023hyperbolic} and have been discussed in Appendix~\ref{app:generator} for a self-consistent narrative.

We introduce a simple tight-binding Hamiltonian with uniform hoppings between nearest neighbours and an on-site potential.
This Hamiltonian can be written in the second quantized form as:
\begin{equation}
    H=\sum_i (Mc_{i,A}^{\dagger} c_{i,A}-Mc_{i,B}^{\dagger}c_{i,B})-t(\sum_{\langle i,j\rangle} c_{i,A}^{\dagger}c_{j,B}+c.c.) 
\end{equation}
Here $c_{i,A(B)}$ represents the annihilation operator at the A(B) sub-lattice of the i-th site in the lattice. 
Likewise, $c_{i,A(B)}^{\dagger}$ represents the creation operator at the A(B) sub-lattice of the i-th site in the lattice.

$M$ defines an on-site potential with opposite signs on either sub-lattice, and $t$ is the strength of the inter-site hopping.
We restrict ourselves to nearest neighbour hoppings to get an analytically tractable model which is also motivated by the circuit QED setups implemented in experimental realizations~\cite{chen2023hyperbolic}. 
We shall show that this model is sufficient to provide an abundance of exceptional contours in the presence of non-Hermitian perturbations.
Just like in Euclidean space, the $k$-space Hamiltonian can be obtained using a U(1) Bloch theory by attaching a phase of $e^{ik_j}$ corresponding to every translation operator $\gamma_j$ applied to perform the hopping. 
This procedure has been detailed in App~\ref{app:generator} to get the Bloch Hamiltonian in the sublattice basis as:
\begin{equation}\label{eq:H0}
    \hat{H}_0=
    \begin{pmatrix}
       M &  -t (1 + e^{ik_1} + e^{ik_2} + e^{ik_3}+ e^{ik_4}) \\
       -t(1+e^{-ik_1}+e^{-ik_2}+e^{-ik_3}+e^{-ik_4}) & -M
    \end{pmatrix},
\end{equation}

For the rest of this work, we set $t=1$.
We can get the tight-binding graphene Hamiltonian from this model by setting $k_3=0$ and $k_4=\pi$.
We note that tuning $k_3$ and $k_4$ provide additional degrees of freedom to the off-diagonal term:

\begin{equation}
h(\mathbf{k}):=1+e^{ik_1}+e^{ik_2}+e^{ik_3}+e^{ik_4}
\end{equation} 
For ease of representation, we will treat the $\mathbf{k}$-vectors as parameters to study the positions of EPs and nodal points in cross-sections of the 4D Brillouin zone.
The simplicity of the Hamiltonian will be helpful to analytically study the onset and behaviour of exceptional contours upon introducing non-Hermiticity.

\begin{figure}
     \includegraphics[width=\textwidth]{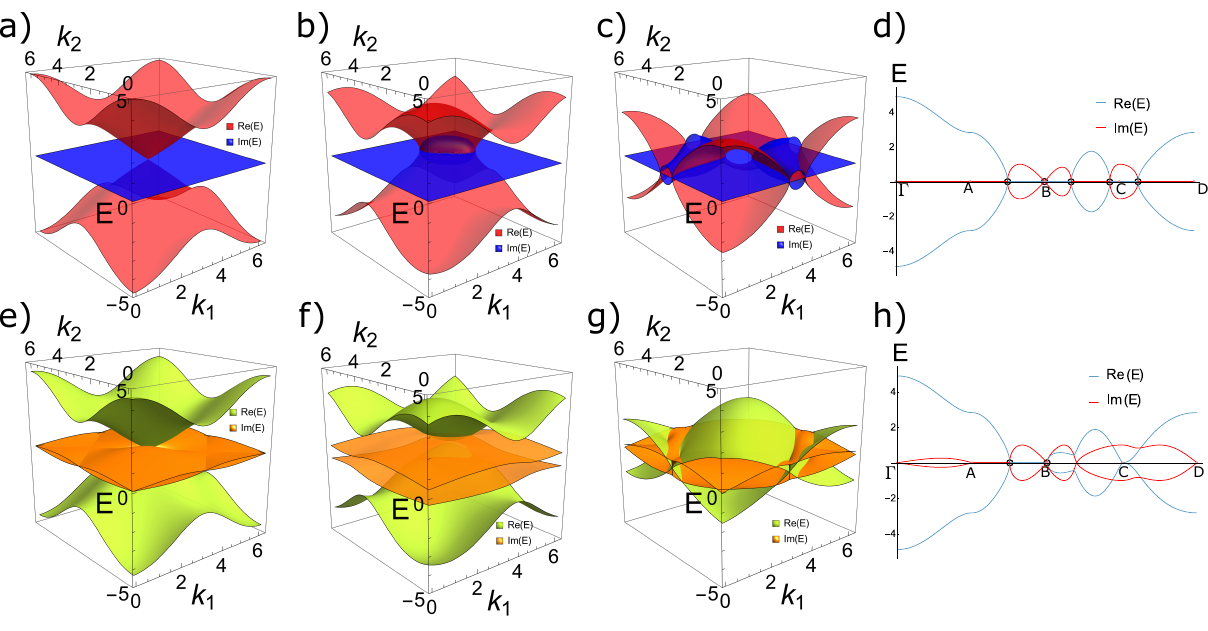}
     \caption{\textbf{Energy spectra for the non-Hermitian cases with gain and loss ($\delta$), and non-reciprocity ($\Gamma$).} (a)-(c) The shape of the nodal contour on the projected $k_1-k_2$ plane with varying $k_3$ and $k_4$ for the case of gain and loss. (a) For $\delta=1$, $k_3=k_4=0$ gives the limiting case of a single band intersection point $\vec{k_0}=(\pi,\pi,0,0)$. The dispersion near this point is quadratic in $\mathbf{k}-\mathbf{k_0}$. (b) A single nodal loop is obtained for $\delta=1, k_3=k_4=\pi$. The spectrum scales as $E\thicksim(|\mathbf{k}-\mathbf{k_0}|)^\frac{1}{2}$ near the nodal contour. (c) Nodal contours obtained for $\delta=1, k_3=7\pi/6, k_4=\pi/4$ (d) The energy spectrum on the same path (but with $\delta=1$) as in Fig.~\ref{fig:Hermspectra} (f). The points where both the energy levels become degenerate are marked with hollow circles, and occur when $|h(\mathbf{k})|=\delta$. (e)-(g) The energy bands and corresponding nodal contours for different choices of parameters $(\Gamma,k_3,k_4)$, i.e., non-reciprocal hopping case. (e) Similar to (a), one obtains a single nodal point where the bands touch. The choice of parameters is $\Gamma=0.5,k_3=\arccos(3/4), k_4=2\pi-k_3$. The dispersion is quadratic in $k$ near the nodal point. (f) Absence of nodal points, with a finite complex gap in the spectrum for parameter values $\Gamma=0.7,k_3=k_4=\pi/2.$ (g) For $\Gamma=0.5,k_3=4\pi/3, k_4=4\pi/5$, open nodal contours are obtained, where the spectrum scales as $E\thicksim(|\mathbf{k}-\mathbf{k_0}|)^\frac{1}{2}$, where $k_0$ is a point on the nodal contour. (h) The energy spectrum is shown on the above-mentioned path, with $\Gamma=1$. There are two nodal points where both the real and imaginary parts of the eigenvalues become equal.}\label{fig:nonhermspectra}
\end{figure}

\subsection{Hermitian spectra}

\begin{figure}
     \includegraphics[width=0.7\textwidth]{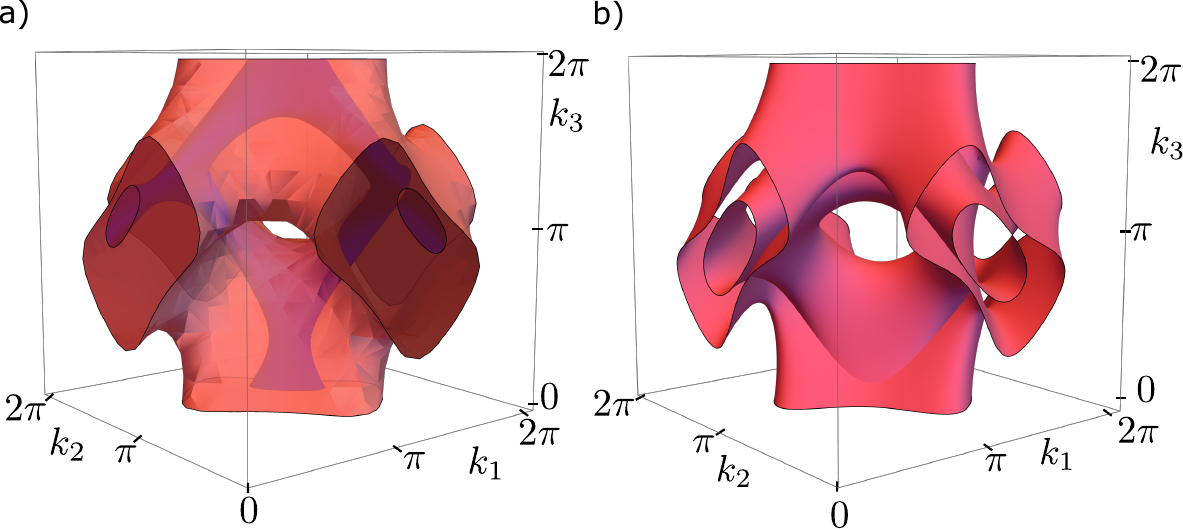}
     \caption{\textbf{Exceptional nodal surfaces for non-Hermitian hyperbolic graphene.} The figures display the nodal surface on the $k_1$,$k_2$,$k_3$ subspace of the Brillouin zone where energy eigenvalues vanish. (a) The nodal surface for the case of on-site gain and loss with $\delta=0.5$. In this case, the condition for a nodal surface is $|h(\mathbf{k})|=|\delta|$. The pink region shows the regions where there exists a node for any value of $k_4\in[0,2\pi]$. The blue subregion is the nodal surface with $k_4=\pi/2$. (b) The nodal surface for the case of non-reciprocal hoppings with $\Gamma=0.5$. The condition for nodes is $h(\mathbf{k})=\pm\Gamma$.
     The nodal surface becomes a two-dimensional surface compared to the three-dimensional volume in (a). This is due to the additional constraint $2i\Gamma \mathrm{Im}(h)=0$ in Eq.~\ref{eq:8}, thus reducing the degrees of freedom of the spectrum.} \label{fig:Nodal}
\end{figure}

We briefly summarize the properties of the Hermitian model in Eq.~\eqref{eq:H0}. 
In the Hermitian regime, the parameters at our disposal are the on-site potential $\pm M$ and the hopping strength $t$ (assumed to be isotropic). 
The energy spectrum is similar to Euclidean graphene and is given by Eq.~\eqref{eq:hermspect}, with the additional freedom furnished by $k_3$ and $k_4$ providing additional tuning parameters on the $k_1$-$k_2$ cross-section. 
\begin{equation}\label{eq:hermspect}
    E_{\pm}=\pm\sqrt{M^2+|h(\mathbf{k})|^2}
\end{equation}
Plotting cross-sections for the energy spectrum with $M=0$ for different choices of $k_3,k_4$ can give nodal points as well as nodal lines as shown in Fig.~\ref{fig:Hermspectra}(a)-(b). 
The nodal contours are surfaces in the Brillouin zone where the bands touch.
For a Hermitian system, the nodal surface corresponds to points of degeneracies where both the eigenvalues are equal. 
Such nodal surfaces have been shown for the ($k_1$,$k_2$,$k_3$) space in Ref.~\cite{chen2023hyperbolic}. 
For a non-Hermitian system, these correspond to exceptional nodal contours (see Fig.~\ref{fig:Nodal}), where the eigenvectors also coalesce.
The locus of the nodal surface is given by $h(\mathbf{k})=0$.
The nodal surface is visualised in Fig.~\ref{fig:Hermspectra}(c), where the surface represents values of $k_1,k_2,k_3$ for which there exists a node for some value of $k_4$. 
Fig.~\ref{fig:Hermspectra}(f) shows the variation of the energy spectrum through different points in the four-dimensional Brillouin zone.
 
A sublattice potential $M$ introduces a gap (equal to 2$M$) in the system, and the linear dispersions at the Dirac points are replaced by quadratic dispersions at the band extrema, as shown in Fig.~\ref{fig:Hermspectra}(d)-(e). 

\subsection{Introducing non-Hermiticity}

We will introduce non-Hermiticity through on-site gain and loss (terms proportional to $\sigma_z$) and non-reciprocal hopping (terms proportional to $\sigma_{x,y}$), by adding terms of the form $i\delta\sigma_z$, $i\Omega\sigma_x$, and $i\Gamma\sigma_y$, to the Hamiltonian $\hat{H}_0$ in Eq.~\eqref{eq:H0}. 
For simplicity, we will consider the non-Hermitian terms to be independent of $k$.
This implies that the non-reciprocity is contained within the unit cell.
We will see how this anisotropy in the non-Hermiticity gives a rich complex energy spectrum with an abundance of EPs. 

Further, the higher dimensionality of the $k$-space endows greater freedom to tune the parameters of the system, allowing us to realise rich non-Hermitian phenomena. 
This freedom will have an even greater significance in obtaining higher-order EPs for tessellations with larger unit cells, such as the four band \{8,4\} model, which will be discussed later.

\subsubsection{On-Site Gain and Loss}

First, we add an on-site gain and loss of strength $\delta$, by adding $i\delta\sigma_z$ to $\hat{H}_0$ in Eq.~\eqref{eq:H0}. 
As a result of the non-Hermiticity, the spectrum is no longer entirely real. 
In fact, the eigenvalues $E_{\pm}=\pm\sqrt{|h(\mathbf{k})|^2-\delta^2} $ will be purely real or purely imaginary.

For a non-zero $M$, the square root has a constant non-zero imaginary part which prohibits any band touchings, even in the individual real and imaginary spectra.
This can be seen by the replacement $\delta\rightarrow\delta-iM$ in the energy dispersion.
The locus of the real and imaginary energies ($E_r$ and $E_i$) is the intersection of three hyperbolic surfaces, given by,

\begin{align}
        E_r E_i&=M\delta,\\
    E_r^2-E_i^2&\leq M^2+25-\delta^2,\\
    E_r^2-E_i^2&\geq M^2-\delta^2.
\end{align}

Clearly, increasing the non-Hermiticity strength $\delta$ will lead to a more prominent imaginary spectrum. 
For $M=0$, we find that either the real or the imaginary part of the eigenvalues will vanish, and we can recover nodal contours where the eigenvalues are zero, with the cross-sections shown in Fig.~\ref{fig:nonhermspectra}(a)-(d) for some choices of $k_3,k_4$. 
Fig.~\ref{fig:Nodal}(a) shows the surface in the $k_1-k_2-k_3$ subspace with $\delta=0.5$ for which a node can be obtained for some value of $k_4$ in the Brillouin zone.

The nodal contours mark the transition from purely real to purely imaginary eigenvalues. 
In Fig.~\ref{fig:nonhermspectra}(d), a one-dimensional contour is parametrised in the Brillouin zone, which shows the appearance of EPs (circled in black). 
Contours connecting these EPs are non-Hermitian Fermi arcs characterised by purely real or imaginary energy eigenvalues\cite{bergholtz2021exceptional}.
While these arcs occur trivially due to the behaviour of the band structure for this case, we will observe their existence even for non-reciprocal hopping in the next section.
In the limit $\delta>5t$, the spectrum becomes purely imaginary, and the bands do not touch in the four-dimensional Brillouin zone. 

\subsubsection{Non-Reciprocal Hopping}

Another way to introduce non-Hermiticity is through non-reciprocal hopping within the unit cell. This can be done by adding terms of the form $i\Omega\sigma_x$ or $i\Gamma\sigma_y$. These terms produce an imaginary $k$-dependent term in the expression for the eigenvalues given by

\begin{equation}\label{eq:7}
    i\Omega\sigma_x: E_\pm=\pm\sqrt{|h|^2+M^2-\Omega^2+2i\Omega \mathrm{Re}(h)},\\
\end{equation}
\begin{equation}\label{eq:8}
    i\Gamma\sigma_y: E_\pm=\pm\sqrt{|h|^2+M^2-\Gamma^2-2i\Gamma \mathrm{Im}(h)}.
\end{equation}

To obtain a node, both the real and imaginary parts in the square root must go to zero.
The shape of the nodal surface is very sensitive to non-reciprocity and changes drastically upon the addition of a small non-reciprocal hopping, as shown in Fig.~\ref{fig:Nodal}(b).
Typical two-dimensional cross-sections of the eigenspectra are shown in Fig.~\ref{fig:nonhermspectra}(e)-(h), with different nodal structures, which could be a nodal point (e), a nodeless cross-section (f), or a nodal contour (g). 
Fig.~\ref{fig:nonhermspectra}(h) shows the variation of the spectrum on a one-dimensional contour in the Brillouin zone. 
The appearance of the nodal EPs is shown in black, where both the real and imaginary parts of the eigenvalues are equal. 
These nodes are connected through non-Hermitian Fermi arcs where the real part of the energy goes to zero, as shown by our choice of contour.
For the case of the $i\Gamma\sigma_y$, the condition for the appearance of nodes is obtained to be

\begin{align*}
    |h|^2=\Gamma^2-M^2, \\
    \Gamma \mathrm{Im}(h)=0.
\end{align*}

Therefore, for non-zero $\Gamma$, we require $h=\pm\sqrt{\Gamma^2-M^2}$. The requirement for the imaginary part to be zero results in lower dimensional nodal surfaces on the $k_1,k_2$ cross-section as shown in Fig.~\ref{fig:Nodal}(b). 
Similarly, the nodal spectrum for non-reciprocity $i\Omega\sigma_x$ is obtained by setting $h=\pm i\sqrt{\Omega^2+M^2}$. 

\section{Diagnostics of Exceptional Contours}\label{sec:exdiag}

\begin{figure}
     \includegraphics[width=0.95\textwidth]{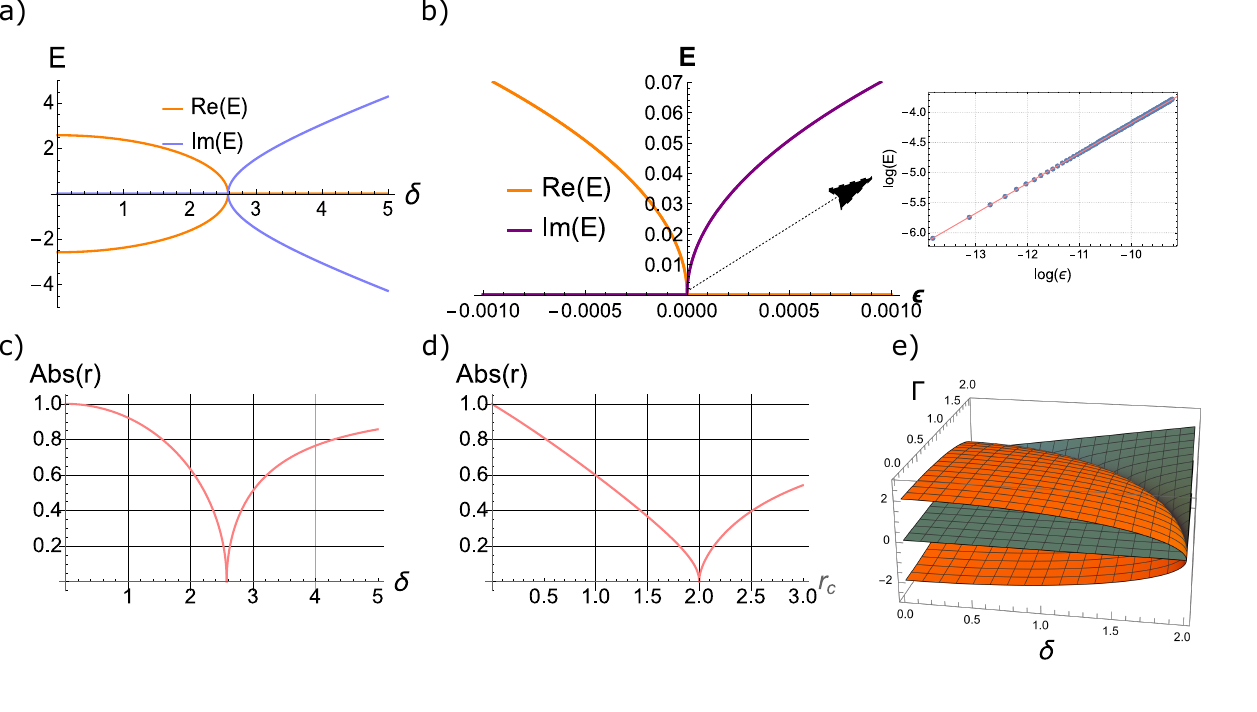} 
     \caption{\textbf{Diagnosing exceptional points using phase rigidity.} (a) The energy spectrum as a function of $\delta$ for $\mathbf{k}=(\pi/2,\pi/3,3\pi/4,\pi)$. The bands touch each other when 
     $\delta=\delta_{EP}=|h(\mathbf{k})|\approx2.58145.$ (b) The scaling of the energy very close to the EP ($\delta=\delta_{EP}+\epsilon$), where $\epsilon$ is a perturbation from the EP). The inset shows the logarithmic scale plot, which has a slope of 0.5, revealing a square root dependence on $\epsilon$ characteristic of second-order EPs. (c) The magnitude of the phase rigidity, $r$, as a function of $\delta$ for the same \textbf{k}. (d) For a system with both non-reciprocity ($\Gamma$) and on-site gain and loss ($\delta$) parametrised by $\delta=r_c\cos\theta, \Gamma=r_c\sin\theta$, the spectrum is independent of $\theta$. For $k_1=4\pi/3, k_2=0, k_3=2\pi/3,k_4=0$, the condition for obtaining a gapless point is $r_c=2$. The phase rigidity is shown to go to zero as a function of $r_c$, with $\theta=\pi/4$. (e) The circular exceptional contour in the $\delta-\Gamma$ parameter space for $\mathbf{k}=(0,0,2\pi/3,4\pi/3)$.} \label{fig:Phase_Rigidity}
\end{figure}

\begin{figure}
    \includegraphics[width=0.6\textwidth]{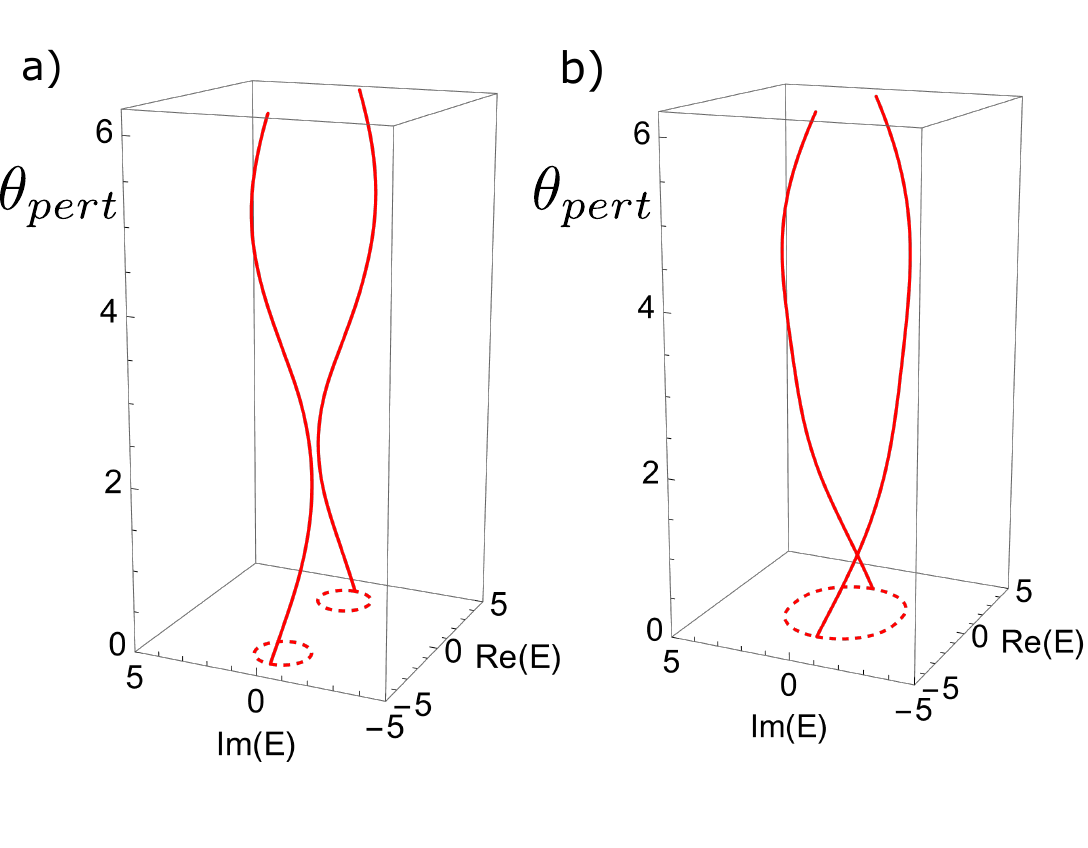}
    \caption{\textbf{Inter-Band winding for the \{10,5\} tessellation.} (a) Illustration of the winding of energy levels for the \{10,5\} tessellation with the on-site gain and loss. For the parameter choice $k_1=2\pi/3,k_2=\pi/3,k_3=0,k_4=0$ and $\delta=2$, the Hamiltonian is not defective, and we get trivial winding of the energy levels with respect to a parametric perturbation $e^{i\theta_{pert}}\sigma_x$ added to $\hat{H}_0$. (b) $\delta=2\sqrt{3}$ makes the Hamiltonian rank-deficit and gives an EP at the chosen $\vec{k}$-point. In this case, we can see the inter-band winding visualised through the exchange of the energy levels in one complete cycle of the perturbation.}
    \label{fig:10_5_Vorticity}
\end{figure}

\subsection{Identifying Exceptional Points}

For non-Hermitian Hamiltonians, the usual orthogonality of eigenvectors is replaced by the weaker condition of \emph{biorthogonality}~\cite{brody2013biorthogonal}. 
This gives us the right and left eigenkets, defined as eigenvectors of $H$ and $H^{\dagger}$ respectively obeying the biorthogonality condition

\begin{equation}\label{eq:biortho}
    \braket{L_i|R_j}=\delta_{ij},
\end{equation}

where $\delta_{ij}$ is the Kronecker delta function, and the eigenvectors are ordered such that $H\ket{R_i}=E_i\ket{R_i}$ and $H^{\dagger}\ket{L_i}=E_i^*\ket{L_i}$.
EPs represent branch points in the spectral manifold and are endowed with topological properties as a consequence of the spectral winding around the branch point~\cite{bergholtz2021exceptional,ding2022non}.
At an EP, two or more eigenvectors coalesce. 
Hence, Eq.~\eqref{eq:biortho} falls through since if $\ket{R_i}=\ket{R_j}$ then their inner products with $\bra{L_i}$ (expected to be 1 and 0 respectively) must also be equal, contrary to the statement of the biorthogonality. 
Thus, at an EP, the state $\ket{R_i}$ cannot be left normalized to unity since $\braket{L_i|R_i}=0$.
The phase rigidity, $r$, measures this biorthogonality to quantify the approach to an EP~\cite{PhysRevE.74.056204,eleuch2016clustering}.
The phase rigidity, $r_j$ for the j-th eigenstate is defined as~\cite{banerjee2022non}

\begin{equation}
    r_j=\frac{\braket{L_j|R_j}}{\braket{R_j|R_j}}.
\end{equation}

For a Hermitian system, $r_j=1$, since $H=H^{\dagger}$, implying the left and right eigenkets are equal.
On the other hand, $r_j=0$ when $\ket{R_j}$ coalesces with another eigenstate at an EP.
Therefore, as we increase the non-Hermiticity, the left and right eigenkets become increasingly orthogonal, leading to a reduction of $r$.
Remarkably, phase rigidity is not just a theoretically defined quantity, but has also been experimentally measured~\cite{tang2020exceptional}. 

EPs lead to degeneracies in the spectrum, which in our case can only occur if the eigenvalues are zero (since the spectrum is symmetric about $E=0$). 
Thus, the nodal surfaces that we obtained earlier correspond to the intersections of the Fermi arcs for the real and imaginary parts of the energy leading to second-order EPs.  
We will subsequently investigate the occurrence of higher-order EPs in a four-band model for the \{8,4\} tessellation where we shall see EPs with all four eigenvectors coalescing.

For second-order EPs, the eigenspectrum shows a square-root scaling, which is displayed in Fig.~\ref{fig:Phase_Rigidity}(a)-(b).
The complex eigenspectrum of the non-Hermitian system allows the definition of the winding of energy eigenvalues, called the vorticity, as a topological invariant~\cite{bergholtz2021exceptional}.
EPs are characterized by fractional values for the vorticity $\nu_n$ of the $n$-th band as a consequence of the coalescing of more than two eigenvectors~\cite{ding2022non}.

\begin{equation}
    \nu_{n}=\frac{1}{2\pi}\oint_{\mathcal{C}}\nabla_{\textbf{k}}\mathrm{arg}(E_n(\mathbf{k}))\cdot d\mathbf{k}.
\end{equation}

On a contour $\mathcal{C}$ encircling the EP, the branch cut causes a net winding of the coalescing eigenvectors. 
As a consequence, the sum of the spectral windings (called the eigenvalue winding number~\cite{ding2022non}) of the eigenvectors is unity ($\Sigma_j \nu_j=1$).

The phase rigidity and vorticity quantify the behaviour of eigenvectors and eigenvalues, respectively, near an EP.
An $n$-th order EP creates a branch point associated with the $n$-th complex root in parameter space.
Assuming the Hamiltonian near an EP to be a small perturbation from the Hamiltonian at the EP, the eigenvalues and eigenvectors can be calculated using a non-Hermitian perturbation theory~\cite{PhysRevLett.125.180403, PhysRevResearch.5.033042}.
This yields the branch point in the eigenvalue dispersion which we quantify using vorticity.
Similarly, the phase rigidity (which vanishes at the EP), can be calculated near the EP to obtain either a $1/n$ or $(n-1)/n$ scaling depending on the order of the EP~\cite{jaiswal2021characterizing}.

For a contour $\mathcal{C}$ encircling the EP, $\nu_{\pm}= 1/2$ for both bands at the second order EP in our model, giving a net winding of 1.
This is shown in terms of a parametrisation of a circular path centred at an EP in terms of the vorticity in Fig.~\ref{fig:10_5_Vorticity}, and leads to an exchange of the eigenvalue and eigenvectors across the branch cut.

\begin{figure}
     \includegraphics[width=0.95\textwidth]{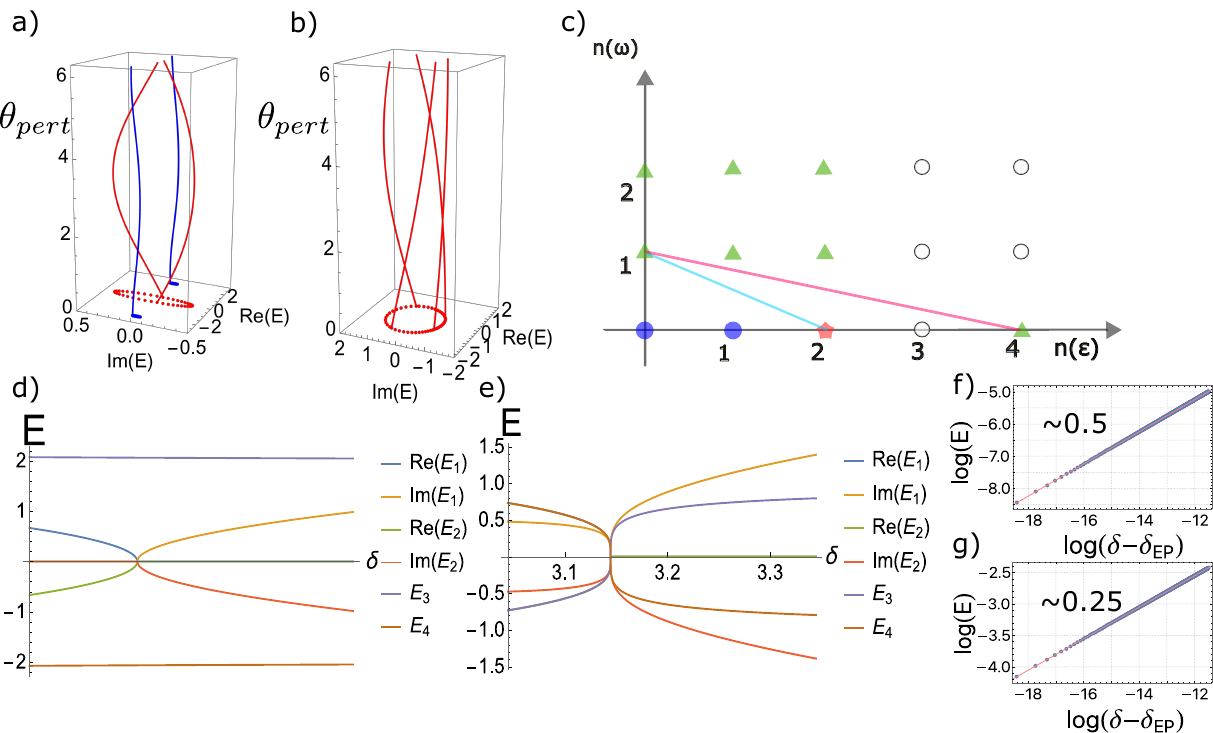}
     \caption{\textbf{Inter-band winding and higher-order EPs in \{8,4\} tessellation.} (a) The eigenvalue winding for the second order EP. The bands in red show an exchange, whereas the bands in blue return to their original values under a cyclic perturbation with $\epsilon=0.07 e^{i\theta_{pert}}$ added as shown in Eq.~\eqref{eq:EP4}. This gives a second-order EP. The parameters taken here are $k_1=\pi/4,k_2=2\pi/3,k_3=3\pi/5,k_4=61\pi/60$ and $\delta\approx2.07923$. (b) The fourth order EP, where $k_1=k_2=0, k_3=k_4=\arccos{(2\sqrt{5}-5)}$ and $\delta\approx 3.14461$. Panel (c) shows the Newton polygon diagram for the \{8,4\} model considered above. The hollow circles show the terms that are absent in the characteristic polynomial. The blue circles are the terms that need to vanish to obtain a second-order EP, whose bounding line is shown in blue. The red point is the additional term that needs to vanish to reach a fourth-order EP, whose bounding line is shown in red. Panels (d) and (e) show the energy spectrum near the EP for the second and fourth-order EPs, respectively. The scaling for the spectra is shown in (f) and (g), respectively. The lowest-order perturbation terms in the spectra have scaling coefficients 0.5 and 0.25, respectively.} 
     \label{fig:8_4_Vorticity}
\end{figure}

\subsection{Higher order Exceptional Points in \{8,4\} Tesselations}\label{sec:higher}

In recent years, higher-order EPs -- where more than two eigenvalues and eigenvectors merge -- have been of significant interest owing to their remarkable properties beyond second-order EPs.
Here, we next show how hyperbolic lattice models allow such higher-order EPs in the presence of non-Hermiticity. 
As an example, we consider the \{8,4\} tessellation, which is part of the \{4$g$, 4\} infinite family of tessellations with genus 2~\cite{boettcher2022crystallography}. 
The unit cell has four sites and the Bolza lattice is an \{8,8\} tessellation, with the discrete Fuchsian symmetry group generated by four elements.
Therefore, this four-band system gives us a simple system to realize second and fourth-order EPs.
We implement a similar nearest neighbour tight-binding model using the translation operators for the \{8,8\} lattice.
Adding a balanced complex gain and loss $\pm i \delta$ on two sites gives the $k$-space Hamiltonian:

\begin{equation}
    H_0(\mathbf{k})=\begin{pmatrix}i\delta & 1+e^{i(k_1-k_2)} & 0 & e^{ik_1}+e^{-ik_4}\\1+e^{i(k_2-k_1)}&0&1+e^{i(k_2-k_3})&0\\0&1+e^{i(k_3-k_2)}&0&1+e^{i(k_3-k_4)}\\e^{-ik_1}+e^{ik_4}&0&1+e^{i(k_4-k_3)}&-i\delta\\ \end{pmatrix}.
\end{equation}

To study the EPs generated in this model, we use the Newton polygon method elaborated in Ref.~\cite{jaiswal2021characterizing}. 
For a contour enclosing an $n$-th order EP, we obtain a mutual winding of $n$ eigenvalues, leading to a cyclic permutation of the eigenvalues across the branch cut.
At the $n$-th order EP, the eigenvalues of the perturbed Hamiltonian show a $1/n$ scaling with the perturbation $\epsilon$.
The Newton polygon method uses this scaling relation in the binomial expansion for the secular equation, given by

\begin{equation}
    P(\omega,\epsilon)=\det(H+\epsilon H_1-\omega\mathcal{I})=\sum_{n(\epsilon),n(\omega)} p_{n(\epsilon),n(\omega)} \epsilon^{n(\epsilon)}\omega^{n(\omega)},
\end{equation}

where $\epsilon H_1$ represents a perturbation of strength $\epsilon$ added to the Hamiltonian $H$ with parameters tuned to give an EP.
The points $(n(\epsilon),n(\omega))$ with $p_{n(\epsilon),n(\omega)}\neq0$ are plotted in $\mathbb{R}^2$, which gives the Newton polygon as the smallest convex polygon containing all the plotted points.
The negative inverse of the slope of a line with all points on or above it gives the lowest order scaling of $\omega$ with $\epsilon$ ($1/n$), and thus the order of the EP ($n$).

We implement this method to study the \{8,4\} lattice.
We begin by introducing a perturbation, $\epsilon$, to the hopping between the first two sites in the unit cell such that the perturbed Hamiltonian becomes

\begin{equation}\label{eq:EP4}
    H(\mathbf{k})=\begin{pmatrix}i\delta & 1+e^{i(k_1-k_2)}+\epsilon & 0 & e^{ik_1}+e^{-ik_4}\\1+e^{i(k_2-k_1)}+\epsilon&0&1+e^{i(k_2-k_3})&0\\0&1+e^{i(k_3-k_2)}&0&1+e^{i(k_3-k_4)}\\e^{-ik_1}+e^{ik_4}&0&1+e^{i(k_4-k_3)}&-i\delta\\ \end{pmatrix}.
\end{equation}

The Newton polygon diagram for our Hamiltonian is shown in Fig.~\ref{fig:8_4_Vorticity} (c). 
Using the analytic form for the characteristic polynomial, we can fine-tune different parameter choices to eliminate the (0,0) and (1,0) points on the diagram to obtain a second-order EP in the system. 
Subsequently, we can remove the (2,0) point to obtain a fourth-order EP. 
The energy spectra for the four bands are shown in Fig.~\ref{fig:8_4_Vorticity}(d)-(e) for the second and fourth-order EPs, respectively. The $E_3,E_4$ bands in Fig.~\ref{fig:8_4_Vorticity}(d) do not mix and remain purely real. The scaling for $E_1,E_2$ bands is shown in Fig.~\ref{fig:8_4_Vorticity}(f), revealing a square root dependence on the parameters, as expected for a second-order EP.

For the fourth-order EP, all four bands participate in the mixing process and scale more steeply as a function of $\delta$. 
The logarithmic scaling shows a $\delta^{1/4}$ dependence on the parameters, as expected for a fourth-order EP. 
The obtained scaling is shown for the spectrum near the EPs in both cases in Fig.~\ref{fig:8_4_Vorticity}(f) and (g), thereby confirming the presence of higher(fourth)-order EPs in this model. 

\begin{figure}
     \includegraphics[width=0.95\textwidth]{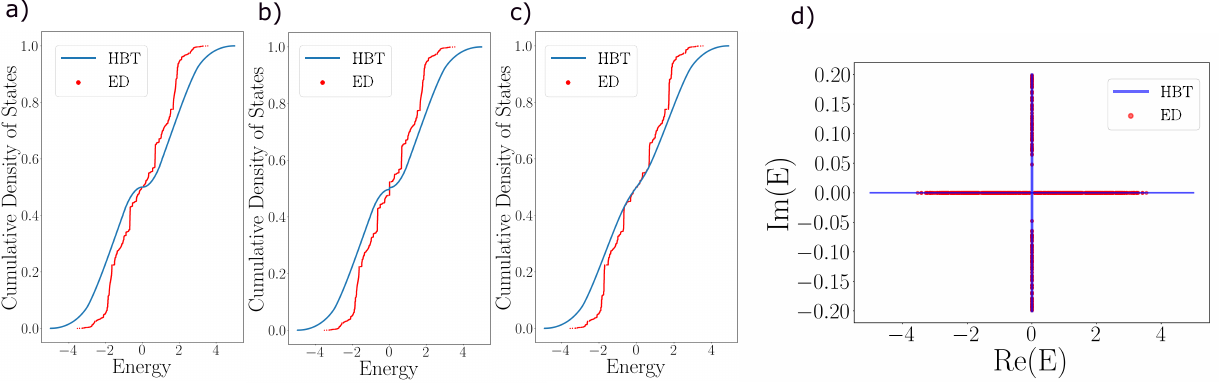}
             \caption{\textbf{Comparing the density of states between hyperbolic band theory (HBT) and exact diagonalization (ED).} (a) The density of states in a cumulative probability distribution (The cumulative probability $p(E)$ is defined as the fraction of states with energy less than or equal to $E$. The cumulative density of states is $p(E)$.) for the Hermitian $k$-space model with periodic boundaries and the open boundary system. 
             (b) The cumulative density of states(p(E)) for the system with $\delta=0.2$. The density is measured with respect to only the real part of the energy, leading to a spike in the probability distribution at $E=0$ corresponding to the spectrum along the imaginary axis as shown in Fig.~\ref{fig:Real_Space}(d).
             (c) The cumulative density of states for the system with non-reciprocity $\Omega=0.2$.
             (d) The complex energy spectra with on-site gain and loss ($\delta=0.2$). The open boundary spectrum is a subset of the periodic boundary condition spectrum.
             \label{fig:Real_Space}.}
\end{figure}

\begin{figure}
     \includegraphics[width=0.95\textwidth]{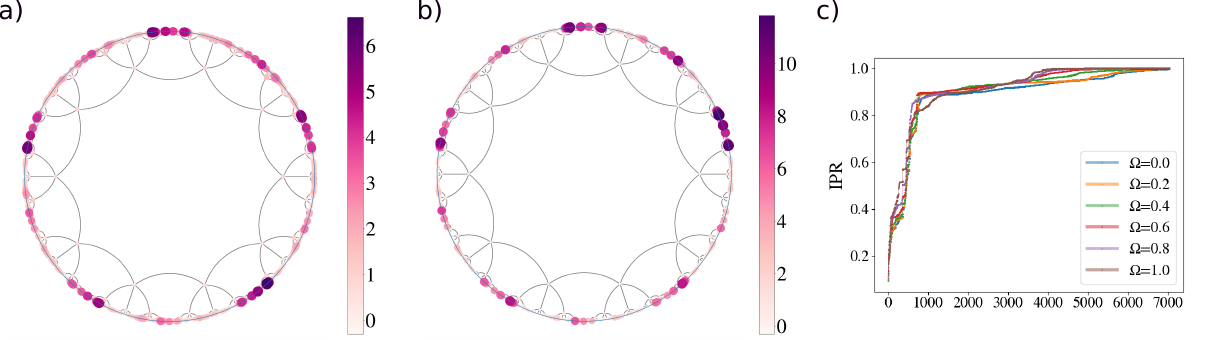}
\caption{\textbf{localization effects with non-reciprocal hopping.} The density of states summed over all the eigenfunctions for the system with two different values of the non-reciprocity parameter (a) $\Omega=0.2$ and (b) $\Omega=0.4$. Although localization can be seen at the outer epochs, the profile is not tenfold rotationally symmetric due to the asymmetry while defining unit cells for non-Hermitian hopping. The extent and magnitude of boundary localization increase with $\Omega$. The asymmetry in the localization is more evident for larger values of $\Omega$. (c) The inverse participation ratio (IPR) for eigenvectors with different values of non-reciprocity. The macroscopic fraction of boundary sites leads to a high IPR even for states of the Hermitian system. The absence of an extensive shift of boundary localised modes with non-reciprocity rules out the possibility of a skin effect in this class of models.} \label{fig:localization}
\end{figure}

\section{Implications in real space: Poincar\'e disk}\label{sec:realspace}

Till now, we have focused on the reciprocal space picture of the exceptional contours. 
Now, we construct a real space system for the \{10,5\} tessellation using the process of circular inversion as outlined in Ref.~\cite{liu2022chern}. 
For the sake of self-consistency, we have included this procedure in Appendix~\ref{app:poinc}.
This method allows the construction of successive epochs of the hyperbolic lattice recursively. 
The number of sites in each successive epoch increases exponentially, and we limit the system size to four epochs (which amounts to 7040 lattice sites). 

Defining the sublattice and unit cells is a recursive variation of the dimer covering problem on the hyperbolic lattice and results in a few dangling sites at the boundary which are not paired into any unit cell. 
This manifests itself as a weak perturbation that breaks the tenfold rotational symmetry of the lattice. 

We numerically compute the density of states with respect to the real part of the energy. Fig.~\ref{fig:Real_Space}(a) shows the comparison of the density of states obtained for the Hermitian system under open boundaries and in the $k$-space. 
The deviation between Bloch theory and exact diagonalization is due to the macroscopic fraction of boundary sites for hyperbolic tessellations as well as the presence of higher-dimensional irreps for the translational symmetry group.
Such deviations between exact diagonalization on finite flakes with open boundaries and band theory have been demonstrated in Ref.~\cite{urwyler2022hyperbolic,chen2023hyperbolic}. 
Recent works have used continued fraction expansions~\cite{mosseri2023density}, supercells~\cite{lenggenhager2023non}, and converging periodic boundary conditions~\cite{lux2023spectral,PhysRevLett.131.176603} to remove spurious contributions from dangling edges and get true bulk spectra from a finite-sized lattice.
Recently, the process of using regular maps to obtain a boundary-removed periodic spectrum has been proposed~\cite{stegmaier2022universality,chen2023hyperbolic} to get better agreements with the $k$-space density of states and remove the effect of the dominating boundary contribution.
However, we will use open boundary spectra to probe localization effects in the non-Hermitian hyperbolic system, since we are also interested in investigating the presence of a non-Hermitian skin effect.

The spectra and density of states for the system with gain and loss are shown in Fig.~\ref{fig:Real_Space}(b)-(d). 
The open boundary spectrum obtained by exact diagonalization(ED), is a subset of the hyperbolic band theory(HBT) spectrum as seen in Fig.~\ref{fig:Real_Space}(b) -- this is consistent with the theorem outlined in Ref.~\cite{okuma2020topological}. 
Whereas the local density of states profile does not show any localization features, the density of states shows a jump at zero energy due to a large number of states with purely imaginary energy. 

Upon adding non-reciprocal hopping, we notice a localization at the outer epoch of the system, as shown in Fig.~\ref{fig:localization}(a)-(b). 
The localization increases with non-reciprocity and has an asymmetric profile due to the method of defining the unit cells, as discussed above. 
We use the inverse participation ratio (IPR) to study this localization by measuring the fraction of wave-function probability at the boundary. 
The IPR for an eigenket $\ket{\psi_n}$, is defined as

\begin{equation}
IPR=\frac{\sum_{r\in \rm{boundary}}|\braket{\mathbf{r}|\psi_n}|^2}{\sum_{r}|\braket{\mathbf{r}|\psi_n}|^2}.
\end{equation}

The numerator calculates the total occupation probability of boundary sites, normalised by the total probability density.
Plotting the IPR for different values of non-reciprocity in Fig.~\ref{fig:localization}(c) reveals that the non-reciprocity does not cause a skin effect in the Hamiltonian and is simply a perturbation to the density profile, enhanced by the microscopic probability density in the wave-function density profile.
In the Hermitian limit under the Bloch ansatz itself, having a macroscopic number of sites at the boundary leads to states with a large IPR, as can be seen from the plot with $\Omega=0.0$ in Fig.~\ref{fig:localization}(c).

\section{Summary and outlook}\label{sec:summary}

In this work, we introduce non-Hermiticity to simple nearest neighbour tight-binding models on hyperbolic lattices to obtain an abundance of exceptional degeneracies.
We assume an automorphic Bloch ansatz for the hyperbolic system to obtain the k-space Hamiltonian in the 4-dimensional Brillouin zone.
The two-band Bloch Hamiltonian for the \{10,5\} serves as a convenient and analytically tractable model to show the exceptional tuning upon the introduction of on-site gain and loss and non-reciprocal hoppings.
Using the phase rigidity and vorticity, we study the behaviour of the complex eigenspectrum near the EPs.

To demonstrate the tuning for higher-order EPs, we use the \{8,4\} lattice with a four-band Hamiltonian. Using the recently proposed method of Newton’s polygons, we can fine-tune the parameters to obtain both second and fourth-order EPs on the \{8,4\} model.

We compare the results from Bloch theory with results from exact diagonalisation for a finite-sized flake with OBC, which shows a deviation in the density of states expected due to the contributions from dangling boundary sites and higher-dimensional irreps of the symmetry group.
A boundary localization is observed for the flake, which is shown to be different from the non-Hermitian skin effect and arises from the non-Hermitian perturbation.

Given the recent experimental realisation of hyperbolic lattices~\cite{kollar2019hyperbolic,chen2023hyperbolic} using circuit quantum electrodynamics and the simulation of non-Hermitian phases using optical and electrical circuit elements~\cite{hodaei2017enhanced,chen2017exceptional,wiersig2020review}, engineering non-Hermitian lattices in hyperbolic geometry is an experimentally viable endeavour.
The abundance of EPs can have potential applications in designing non-Hermitian sensors operating on the scaling of eigenvalues near EPs~\cite{PhysRevLett.112.203901, wiersig2020review,de2022non}.
We hope our work motivates further theoretical and experimental explorations of non-Hermitian hyperbolic matter. 

\textit{Note Added:} During the final stages of this work, we came across a complementary preprint, which studies the skin effect in hyperbolic topological lattice models~\cite{sun2023hybrid}.

\section*{Acknowledgments}
We thank Adhip Agarwala and Vijay Shenoy for illuminating discussions. N.C. acknowledges a fellowship from the Kishore Vaigyanik Protsahan Yojana (KVPY). A.N. is supported by the Indian Institute of Science.

\appendix

\section{Poincaré disk projection}\label{app:poinc}
Projection models such as the Poincaré disk are used to conveniently represent hyperbolic geometries.
Such models are constructed using distance-preserving maps (isometries) from the hyperbolic space to the representation space, in this case, $\mathbb{D}$.
The Poincaré disk projection is frequently used to describe hyperbolic lattices since it allows a compact description of the infinite hyperbolic plane.
Distances in the hyperbolic space are calculated using the Riemannian metric

\begin{equation}
    ds^2=\frac{4|dz|^2}{(1-|z|^2)^2}.
\end{equation}

where $z=x+iy$.
The translation operators for the Poincaré disk are isometries belonging to the Fuchsian group with their operation defined as

\begin{equation}\label{eq:mobius}
    \begin{pmatrix} \alpha & \beta\\
    \gamma & \delta
    \end{pmatrix}z:=\frac{\alpha z+\beta}{\gamma z+\delta},
\end{equation}

where $\alpha, \beta, \gamma, \delta \in \mathbb{R}$ such that $\alpha\delta-\beta\gamma>0$
Here, we discuss the procedure to construct the Poincaré disk projection for a hyperbolic tessellation following the method outlined in Ref.~\cite{liu2022chern}.
This method uses recursive circular inversion to generate outer epochs using inner epochs. 
The choice of \{$p$,$q$\} uniquely fixes the positions of all the sites in the disk $\mathbb{D}$ up to a global rotation of the system.

\begin{enumerate}
\item Taking the origin as the centre of the tessellation and the first site to lie along the positive $x$-axis, the construction of the first epoch can be trivially carried out as

\begin{equation}
    z_n=d_{pq}e^{2\pi i n/p},\quad d_{pq}=\sqrt{\frac{\cot{\pi/q}-\tan{\pi/p}}{\cot{\pi/q}+\tan{\pi/p}}},
\end{equation}

for $n=1,2...p$ and the positions of the sites are stored as complex numbers.

\item The circular inversion of point A with respect to a circle centred at O with radius $r$ is defined as the point A' such that OA and OA' are collinear with OA.OA'=$r^2$.

\item We take pairs of neighbouring sites, say A and B and invert them with respect to $\mathbb{D}$ to get A' and B'.

\item Inverting the remaining sites in the regular polygon containing A and B with respect to the circle circumscribing ABB'A' to get the sites corresponding to the next epoch.

\item This process is repeated for pairs of neighboring sites to construct subsequent epochs.
\end{enumerate}

\section{Algebra of the generators of hyperbolic translations}\label{app:generator}
The group structure of translational symmetry operators plays a crucial role in the band theoretic description of systems. 
Since translations commute in Euclidean space, they form an Abelian group which only permits 1D irreps. 
Bloch’s theorem states that systems (Hamiltonians) with translational symmetries will have eigenvectors that transform as the 1D irreps, and therefore pick up a U(1) phase under the action of the operator.

Since hyperbolic translation operators, represented in Eq.~\eqref{eq:mobius} do not commute, the translational group will, in general, have higher dimensional irreps. 
This implies that the U(1) Bloch theorem cannot be applied to arbitrary eigenvectors for Hamiltonians with hyperbolic translational symmetry. 
For Euclidean geometries, the group of lattice translations forms a normal subgroup for the $d$-dimensional manifold ($\mathbb{R}^d$) whose quotient group gives the Brillouin zone ($\mathbb{T}^d$) – a $d$-dimensional torus.

On the other hand, for a hyperbolic surface, the quotient group of the hyperbolic manifold under the non-Abelian Fuchsian translational symmetry group gives a higher genus Riemann surface. 
This leads to a higher genus representation for the hyperbolic Brillouin zone, which emerges as the Jacobian of the Riemann surface~\cite{maciejko2021hyperbolic}.

Translational symmetry operators for hyperbolic lattices are represented as:

\begin{equation}\label{eq:symmmobius}
    \begin{pmatrix} \alpha & \beta\\
    \beta^* & \alpha^*
    \end{pmatrix}z:=\frac{\alpha z+\beta}{\beta^* z+\alpha^*},
\end{equation}

with $|\alpha|^2-|\beta|^2=1$.

The expressions for the nearest neighbour translational group operators have been discussed in Ref.~\cite{chen2023hyperbolic}. For our choice of lattice orientation, they take the form

\begin{equation}
    \gamma_n=\frac{1}{\sqrt{1-\sigma^2}}\begin{pmatrix}
        1 & \sigma e^{2\pi (n-1/2) i/p}\\
        \sigma e^{-2\pi (n-1/2) i/p} & 1
    \end{pmatrix},
\end{equation}

where $\sigma=\sqrt{\frac{\cos{2\pi/p}+\cos{2\pi/q}}{1+\cos{2\pi/q}}}$ and $n=1,2,..., 5$.
These are not all independent since $\gamma_5=-\gamma_1^{-1}\gamma_2\gamma_3^{-1}\gamma_4$~\cite{chen2023hyperbolic}. 
Thus, we get four independent generators for the discrete translational symmetry group of the \{10,5\} tessellation, with the action of each translation $\gamma_n$ associated with a phase addition $e^{i\phi_n}$ to the wave function.
Ref.~\cite{chen2023hyperbolic} found that the five nearest neighbouring sites are identified by the action of $\mathbb{I}, \kappa_1=\gamma_1\gamma_2^{-1}, \kappa_2=\gamma_2\gamma_3^{-1}, \kappa_3=\gamma_1\gamma_4\gamma_3^{-1}, \kappa_4=\gamma_2\gamma_3^{-1}\gamma_4\gamma_3^{-1}$.
We can use the independent operators $\kappa_1,\kappa_2,\kappa_3,\kappa_4$ as generators of the translation group, with the associated crystal momenta $k_n$ such that the tight-binding Hamiltonian (Eq.[2] in the manuscript) gives the Bloch Hamiltonian

\begin{equation}
    H(\mathbf{k})=\begin{pmatrix}
        M & t(1+e^{ik_1}+e^{ik_2}+e^{ik_3}+e^{ik_4})\\
        t(1+e^{-ik_1}+e^{-ik_2}+e^{-ik_3}+e^{-ik_4}) & -M
    \end{pmatrix}.
\end{equation}

It is important to remember that since we are applying automorphic Bloch theory, we attach a phase of $k_n$ upon the action of $\kappa_n$.
Therefore, the overall phase obtained in a translation will be independent of the order in which the translations have been applied -- it acts as if the translations commute, which is true for 1D irreps.
Note that the $\{k_n\}$ are related to the original $\{\phi_n\}$ as~\cite{chen2023hyperbolic}

\begin{equation}
\mathbf{k}=\begin{pmatrix}
    1 & -1 & 0 & 0\\
    0 & 1 & -1 & 0\\
    1 & 0 & -1 & 1\\
    0 & 1 & -2 & 1
\end{pmatrix}\mathbf{\phi}.
\end{equation}

\bibliography{bibliography1}
\end{document}